\documentclass[aps,pra,twocolumn,showpacs,groupedaddress,amsmath,amssymb]{revtex4}

\usepackage{graphicx}

\usepackage{color}
\usepackage{comment}
\usepackage{bm}
\usepackage{amsmath}
\usepackage{latexsym}
\newcommand{\tp}[1]{\textcolor{red}{#1}}

\definecolor{dgreen}{rgb}{0,0.5,0}

\begin{document}

\title{Giant Non-reciprocity Near Exceptional Point Degeneracies}

\author{Roney Thomas$^*$, Huanan Li\footnote{The first two authors contributed equally to this work}, F. M. Ellis, $\&$ Tsampikos Kottos}
\address{Department of Physics, Wesleyan University, Middletown, CT-06459, USA}

\begin{abstract}
We show that gyrotropic structures with balanced gain and loss that respect anti-linear symmetries exhibit a giant non-reciprocity at the so-called 
exact phase where the eigenfrequencies of the isolated non-Hermitian set-up are real. The effect occurs in a parameter domain near an exceptional point (EP) 
degeneracy, where mode-orthogonality collapses. The theoretical predictions are confirmed numerically in the microwave domain, where a non-reciprocal 
transport above $90$dB is demonstrated, and are further verified using lump-circuitry modeling. The analysis allows us to speculate the universal nature of 
the phenomenon for any wave system where EP and gyrotropy can co-exist.
\end{abstract}
\maketitle

\section{Introduction}

Exceptional points (EP) are non-Hermitian degeneracies where both eigenvalues and eigenvectors coalesce \cite{K80}. Originally 
treated as mathematical curiosities \cite{B04,BB98,M11,H12}, these degeneracies have been now recognized as a source of many counter-intuitive phenomena, 
some of which can be exploited for technological purposes. Examples include loss-induced transparency \cite{Guo}, unidirectional invisibility \cite{LREKCC11,
FXFLOACS13,GSDMVASC09}, lasing mode selection \cite{HMHCK14}, lasing revivals and suppression \cite{PORYLMBNY14}, directional lasing \cite{lan1}, 
hypersensitive sensors \cite{lan2} etc. 

The wealth of these results and the demonstrated capability of the researchers to utilize EPs in order to design novel devices, motivated us here to 
employ them for the realization of a new class of photonic isolators and circulators with an extraordinary (giant) non-reciprocal transport.
The proposed structures
are linear, they involve gyrotropic elements, and they operate in a parameter domain, near an EP degeneracy, where they are stable i.e. the eigenfrequencies of the 
associated isolated set-up are real \cite{RLKKKV12,note0}. The latter two ``conflicting" requirements can be satisfied simultaneously by a class of non-Hermitian 
systems which involve balanced gain and loss mechanisms and which respect antilinear symmetries \cite{BB98}. The parameter domain for which the eigenfrequencies 
are real (stable domain) is known as exact phase while the domain for which the spectrum consists of conjugate pairs of complex eigenvalues (unstable domain) is 
known as broken symmetry phase. The transition between these two phases occurs via an EP \cite{BB98}. A prominent example of such antilinear systems are structures 
with parity-time (${\cal PT}$) symmetry \cite{RMGCSK10,Guo,Konotop,FXFLOACS13,GSDMVASC09, CJHYWJLWX14,POLMGLFNBY14,HMHCK14,LREKCC11,L09a,ZCFK10,
RKGC10,BFBRCEK13,RLKKKV12,MGCM08,PORYLMBNY14,NBRMCKF14,LTEK16}.  

In this paper we demonstrate the EP-induced giant non-reciprocity in the microwave domain and establish its universal nature by evincing it in a seemingly different 
framework of lumped electronic circuitry. The frequency for which the giant non-reciprocity occurs depends on the values of the gain and loss parameter and the applied 
magnetic field. Our approach provides several degrees of reconfigurability, thus constituting an alternative pathway \cite{CJHYWJLWX14,POLMGLFNBY14,RKGC10,BFBRCEK13,NBRMCKF14,FHWWCYL11,SDBB94,FSK05,YF09,MDE14,DBB12} towards enhancing non-reciprocal 
wave transport.

The structure of the paper is as follows. In the next section II we present the photonic structure. Specifically in subsection II.A we analyze the "evolution" of the eigenfrequencies
of the isolated set-up as a function of the gain and loss parameter while in subsection II.B we present the numerical results for the scattering properties of this
structure. In subsection II.C we analyze theoretically using coupled mode theory the transport characteristics of the photonic structure and compare our theoretical
results for the non-reciprocal transmission with the numerical data. At section III we analyze numerically a user-friendly model of coupled LRC lump circuits
and show that also this system demonstrate the same strong non-reciprocal transport. Our conclusions are given at the section IV.

\begin{figure}[h!]
\centering
\hspace*{-0cm}\includegraphics*[width=8cm]{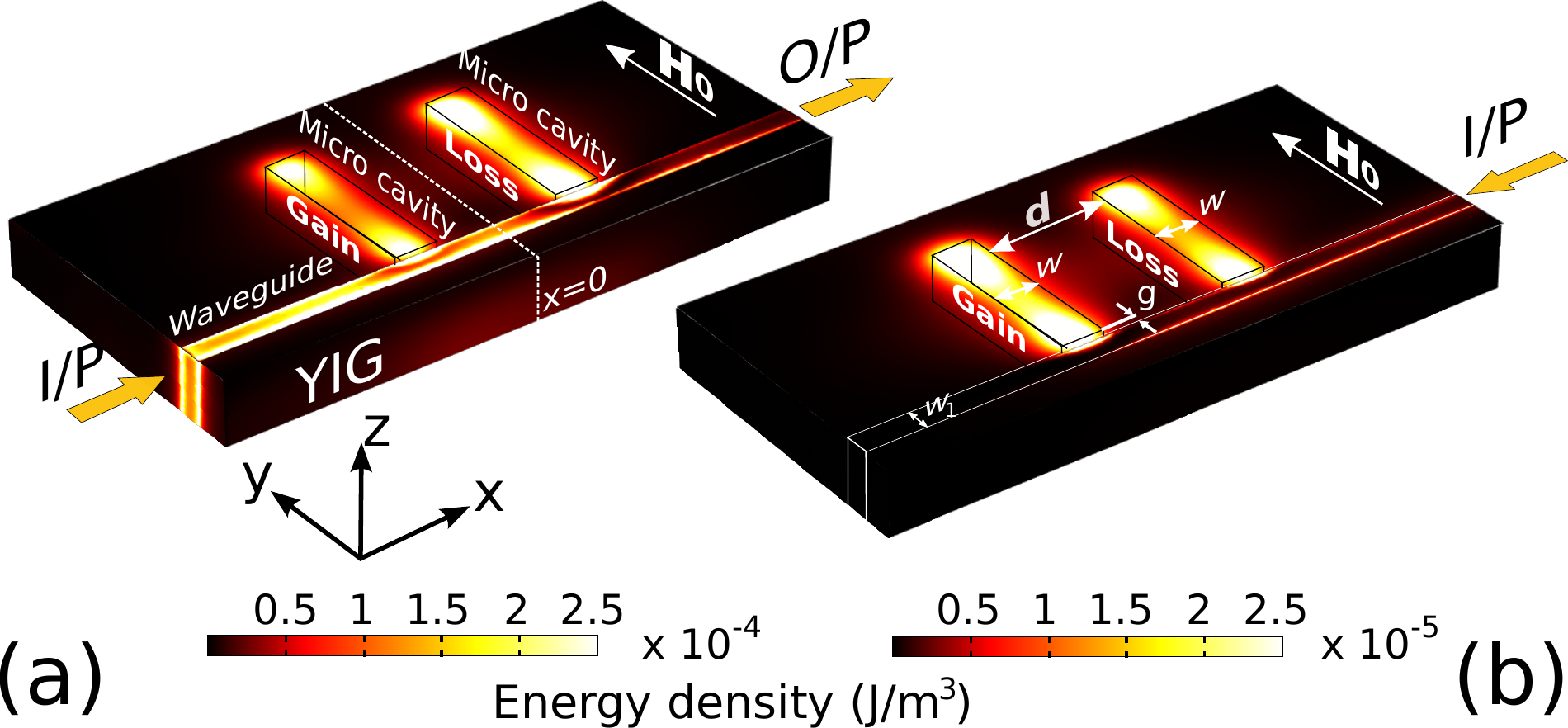}
\caption{Schematic of the photonic structure: two half-wave microstrip resonators are end-coupled to a bus waveguide. A uniformly 
distributed gain or loss material property augments the region beneath each of the resonators within the YIG-substrate. The substrate is 
exposed to an external bias field, H$_0$, in the $y$-direction. For an appropriate value of the gain and loss parameter $\gamma$
the transmission in the forward direction take values of order of unity (a) while it is essentially zero in the backward direction (b). 
}
\label{fig1}
\end{figure}

\section{Photonic structure} 

We consider the structure shown in Fig. \ref{fig1}. It consists of a parallel pair of half-wave microstrip resonators (dimer) 
end-coupled to a bus waveguide as schematically illustrated in Fig.~\ref{fig1}. The microstrip resonator dimer and the waveguide are situated on top 
of an 8.75 mm thick ferrite substrate with a ground plane on the lower surface. The length, $l$, of each microstrip is 24.5 mm, which 
corresponds to an uncoupled half-wave resonance of approximately 1.24 GHz. The widths $w$ and $w_1$ of the microstrips and bus 
waveguide are set at 3.5 mm and 3.0 mm respectively, the latter matching the 56 Ohms impedance of the input bus ports. The 
distance, $d$, between the two microstrip resonators, is set to 20 mm and the end-coupled gap $g$ between the microstrip resonator dimer 
and the bus waveguide is 0.5 mm. All metallic surface structures are defined as zero-thickness, perfect electric conductors. A relative dielectric 
permittivity $\epsilon_r=15$ is used for the ferrite substrate~\cite{Shvets_14,Song_Wu10} matching Yttrium Iron Garnet (YIG). In all our  
simulations, gain and loss are confined to the spatial domain beneath each of the microstrip resonators and implemented by introducing an 
imaginary part of the complex permittivity defined as $\epsilon_r = 15(1\pm i \gamma$), wherein $\gamma$ denotes the gain and loss 
parameter. A practical way of implementing loss or amplification (gain) locally (within the microcavities) can be achieved electronically via 
discrete electronic (loss) or gain devices such as a (resistor), transistor, or tunnel diode~\cite{Fred_12,Matolak_12}. 

A static magnetic bias field, $H_0$, is applied along the $y$-direction through the substrate material having an anisotropic 
magnetic permeability tensor, $\hat{\mu}$, given by:
\begin{equation}
\hat{\mu}=\mu_0\begin{bmatrix}
	\mu_r & 0 &i\kappa_r \\
	0 & 1 & 0 \\
	-i\kappa_r & 0 & \mu_r\\
\end{bmatrix};\, \mu_r = 1+\kappa_r; \, \kappa_r = \frac{\omega\omega_m}{\omega_0^2-\omega^2}, 
\label{const}
\end{equation}
where $\omega_0 = \mu_0 \gamma_e H_0, \quad \omega_m = \mu_0 \gamma_e M_s$. Here, $\mu_0$ and $\omega$ corresponds to the 
permeability of free space and angular frequency, $\omega_0$ corresponds to the precession frequency of an 
electron in the applied magnetic field bias, $H_0=1.273 \times 10^5$ A/m, $\omega_m$ is the electron Larmor frequency at the saturation 
magnetization, $M_s=1.393\times 10^5$ A/m of the ferrite medium, and $\gamma_e$ is the gyromagnetic constant of 1.76 $\times$10$^{11}$ 
rad/sT.

The whole structure satisfies a combined mirror-time symmetry with respect to the $yz$-plane at $x=0$. The mirror-symmetry operator ${\cal M}$ 
is linear and it is associated with a reflection $(x,y,z) \rightarrow (-x,y,z)$ around the origin. The time reversal operator ${\cal T}$ is antilinear and it 
is associated with a complex conjugation together with a simultaneous inversion of  the magnetic field vectors, ${\vec H}_0 \rightarrow -{\vec H}_0$. 
The mirror-time reversal symmetry belongs to the class of anti-linear symmetries, part of which is also the parity-time (${\cal PT}$) symmetry. 
In order to stress this similarity (x-axis parity and to be in direct contact with the vast community that studies transport of ${\cal PT}$-symmetric systems), we 
will abbreviate below the mirror-time reversal symmetry with the letters ${\cal {\tilde P}T}$.

Below we first analyze the parametric evolution of the eigen-frequencies versus the gain and loss parameter of the 
two microstrip system in the absence of the bus waveguide. We refer to this as the ``{\it isolated}" set-up. Its scattering analogue is constructed by 
passing the bus waveguide near one end of the micro-strip pair (see Fig. \ref{fig1}). We refer to this as the ``{\it scattering}" set-up.

The electromagnetic propagation is described by the Maxwell's equations 
\begin{equation}
\vec\nabla \times \vec{E}=i{\omega\over c} {\hat \mu} {\vec H}; \quad
\vec\nabla \times \vec{H}=-i{\omega\over c} {\hat \epsilon} {\vec E} 
\end{equation}
where $\vec{E}$ is the electric $\vec{H}$ is the magnetic field. These equations supplemented by Eqs. (\ref{const}) together with the appropriate 
boundaries dictated by our design of Fig. \ref{fig1} describe the wave propagation from the structure.
The latter is simulated 
with COMSOL's 3D-finite element electromagnetic (FEEM) numerical software~\cite{comsol}.
For accuracy of the numerical results, each domain of the structure comprised of fine mesh 
element sizes of $\approx \lambda_m$/13 within the substrate region and $\approx \lambda_m$/8 for the surrounding air regime, where 
$\lambda_m$ is the wavelength inside the medium. 

\subsection {Isolated set-up}
 
\begin{figure}[h!]
\centering
\hspace*{0cm}\includegraphics*[width=9cm]{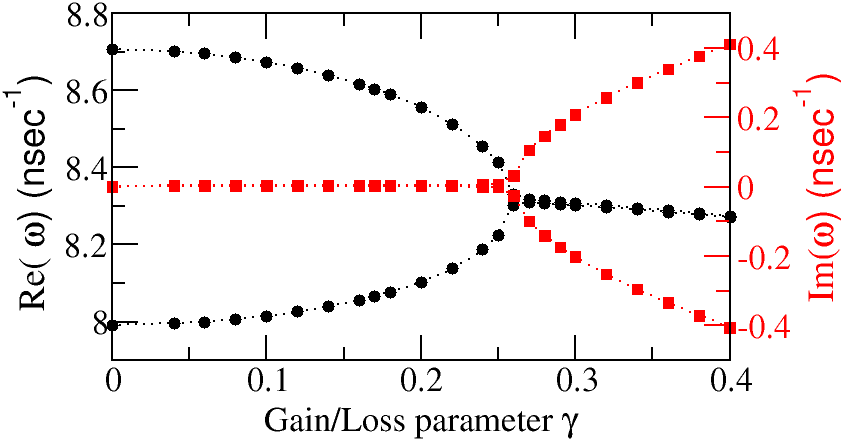}
\caption{Parametric evolution of the real and the imaginary parts of the eigen-frequencies vs $\gamma$ for the isolated set-up of Fig. \ref{fig1}.
A uniform magnetic field $H_0$ is imposed on the substrate. At $\gamma=0$ 
we have a non-zero imaginary part due to leakage from the cavities. At $\gamma=\tp{\gamma_{\cal {\tilde P}T}}\approx 0.26$ an EP 
degeneracy occurs.
}
\label{fig2}
\end{figure}

We investigate the ${\cal M T}$-symmetry phase transition for the isolated set-up of Fig. \ref{fig1} using COMSOL's eigenfrequency simulation. 
When $\gamma=0$, the coupled microstrip resonators support two low-order resonant modes, which have a symmetric (lower frequency $\omega_s$) 
and an antisymmetric (higher frequency $\omega_a$) configuration. For $\gamma=0$ the associated eigenfrequencies have the same imaginary 
value ${\cal I}m\{\omega\}=\eta$ resulting from weak coupling to the perfectly absorbing ends. As $\gamma$ increases the real part of the 
eigenfrequencies of the modes changes (see Fig.~\ref{fig2}) while the associated imaginary part remains the same \cite{note1}. In this domain 
({\it exact phase}) \cite{BB98}, the associated eigenmodes respect the \tp{${\cal {\tilde P}T}$} symmetry. At a critical value of the gain and loss parameter 
$\tp{\gamma_{\cal {\tilde P}T}} \approx 0.26$, the eigenvalues and eigenvectors coalesce and the system experience an EP degeneracy. At the {\it broken phase} 
corresponding to $\gamma>\tp{\gamma_{\cal {\tilde P}T}}$ the real part of the eigenfrequencies remain degenerate while the imaginary 
part bifurcates into two values. We refer to this transition as a {\it spontaneous \tp{${\cal {\tilde P}T}$}-symmetric phase transition}.
The value of \tp{$\gamma_{\cal {\tilde P}T}$} depends on the value (and spatial domain) of the applied magnetic field $H_0$.

\subsection {Scattering set-up} 

\begin{figure}[h!]
\centering
\hspace*{-0cm}\includegraphics*[width=8cm]{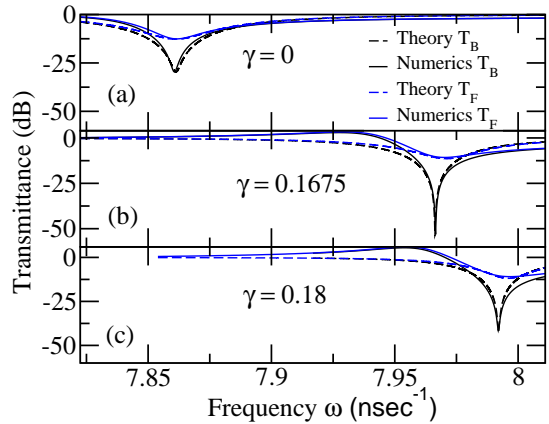}
\caption{Three representative cases of non-reciprocal transport: (a) $\gamma=0$ where $NR=17.5dB$; (b) $\gamma
=0.1675$ where $NR=42.6dB$; and (c) $\gamma=0.18$ where $NR=30.4dB$. The maximum non-reciprocity 
is observed in the domain around $\omega_s$ and it is non-monotonic with respect to $\gamma$.}
\label{fig3}
\end{figure}

Next we proceed with the analysis of the transmission properties of the scattering set-up of Fig. \ref{fig1}. Forward (FWD), 
or backward (BWD) propagation of radiation is defined in the context of the 56 Ohm ports, impedance matched to the transverse electromagnetic (TEM) 
modes from the left and right ends of the bus waveguide shown schematically in Fig.~\ref{fig1}. Our analysis will concentrate on $\gamma$-values for 
which the system is in the exact phase i.e. $\gamma \leq \tp{\gamma_{\cal {\tilde P}T}}$. To quantify the dependence of the non-reciprocal 
effect, we introduce the nonreciprocity parameter $\mathrm{NR}$ (measured in dB), 
\begin{equation}
NR\left(\gamma\right)
= 10 \times {\rm max}_{\omega}\left\{\left|\log_{10}{T_{\rm {B}}\over T_{\rm F}}\right|\right\},
\label{NR}
\end{equation}
where $T_{\mathrm{F}}$ and $T_{\mathrm{B}}$ are the transmittances obtained  for the FWD and BWD cases, respectively. Our numerical investigation 
indicates that the maximum values of $NR$ are achieved in the proximity of the symmetric resonant frequency $\omega_s$. We will therefore focus 
on this frequency domain. In Figs. \ref{fig3}a-c we show some typical transmission spectra for $\gamma$ 
= 0, 0.1675 and 0.18, respectively. Note that at $\omega\approx \omega_s$ the BWD transmittance $T_{\mathrm{B}}$ becomes 
essentially zero while $T_{\mathrm{F}} = {\cal O}(1)$. Specifically for $\gamma=0$ (see Fig. \ref{fig3}a) a non-
reciprocal transmission at $\omega_s$ can be as high as $18 dB$. A higher degree of non-reciprocity $NR=42.6dB$ occurs 
for $\gamma=0.1675$ (see Fig. \ref{fig3}b). However further increase of the gain and loss parameter i.e. $\gamma=0.18$ 
leads to a decrease of non-reciprocity to $NR\approx30.4dB$ (see Fig. \ref{fig3}c). 

The simulation results for $NR(\gamma)$ and its giant enhancement at some critical gain and loss value $\gamma_{\rm NR}$ is reported 
as the solid circles part of Fig. \ref{fig4}b where we show the degree of non-reciprocity $NR$ versus $\gamma$. The non-monotonic behavior 
of $NR$, and the associated maxima, constitute the main result of our study, theoretically discussed in the next section.

\subsection{Theoretical Analysis} 

\begin{figure}
\centering
\hspace*{-0cm}\includegraphics*[width=9cm]{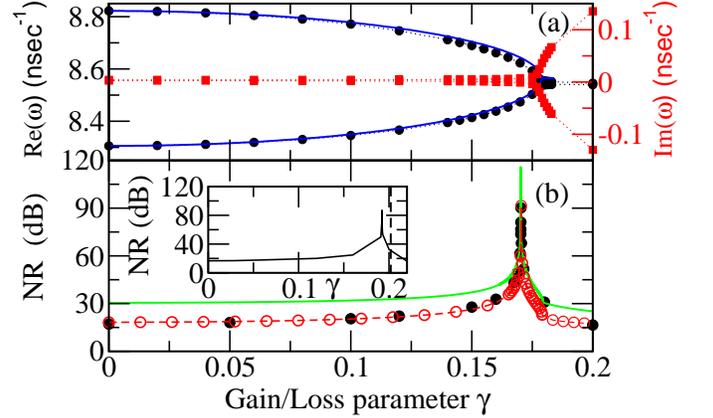}
\caption{(a) We show the dependence of simulated resonant modes (${\cal R}e(\omega), \bullet$ / ${\cal I}m (\omega)$, \tp{$\small \blacksquare$}) 
on the gain/loss parameter $\gamma$ for the set-up with $H_0=0$ only in the domain between the two micro-cavities. A fitting using Eq. (\ref{rfreq}) 
(solid line) gives $\omega_0\approx8.545$, $\Omega_0\approx0.2576$ and $\rho\approx 1.445$ (all measured in nsec$^{-1}$) corresponding to 
$\tp{\gamma_{\cal {\tilde P}T}^0}\approx0.178$. (b) Non-reciprocity (NR) obtained by calculating the difference between the FWD and BWD transmittance $T$ from 
the simulations ($\bullet$) and from the theoretical expressions Eqs. (\ref{TF}) ($\tp{\circ}$). The green line is obtained using Eq. (\ref{NRapprox}). 
The inset shows the analogous simulated NR for a 15\% reduction in the bias field. The vertical dashed line indicates the position of the EP in this case.
}
\label{fig4}
\end{figure}

The behavior of $NR(\gamma)$ seen in Fig. \ref{fig4}b can be understood within the framework of temporal 
coupled-mode theory \cite{H84}. Our calculation scheme breaks down the effect of the magnetic field into two parts. First we consider the 
effect to the resonant frequency of the individual resonators separately (for $\gamma=0$) in a magnetic substrate. For our applied field $H_0$ 
it can be directly estimated from Fig. \ref{fig2} to be $\omega_0\approx 8.2938$ ns$^{-1}$. Next we add the effect of gain and loss $\gamma$ 
in each of these resonances which are now considered as a two level system and coupled via a non-magnetic substrate with a 
coupling constant $\Omega_0$ (i.e. evaluated with $H_0=0$). This is estimated, to a good approximation, from the eigenmode analysis 
of the isolated set-up with $H_0=0$  only in the domain between the two resonators (see Fig. \ref{fig4}a), and is found to be $\Omega_0
\approx 0.2576$ ns$^{-1}$. The resulting symmetric $\omega_s^{(0)}$ and antisymmetric $\omega_a^{(0)}$ 
resonant modes of the isolated composite structure is then:
\begin{equation}
\label{rfreq}
\omega_{s/a}^{(0)}=\omega_0 \mp \sqrt{\Omega_0^2-(\rho\gamma)^2}
\end{equation}
where $\rho\approx 1.445$ ns$^{-1}$ is a scaling parameter that is extracted from the analysis of the isolated set-up of Fig. \ref{fig4}a. 
For this set up, the EP is $\tp{\gamma_{\cal {\tilde P}T}^{0}}=\Omega_0/\rho\approx 0.178. $ 

The second part of our analysis considers the consequences of the magnetic field in the coupling between $\omega_{s/a}^{(0)}$. Specifically, 
we consider that the resonances ($\omega_{s/a}^{(0)}$) are coupled via the magnetized substrate between the two microstrip cavities and indirectly 
via the presence of the bus wave-fields. In general, this additional coupling constant $\lambda$ is a function of the geometric properties of the 
two stripline resonators, the applied magnetic field, $H_0$, and the wavenumber $k_x$ of the bus field. Based on symmetry considerations 
\cite{Shvets_14} we have that up to a linear approximation, $\lambda=\lambda_0+ \imath\left(b_0k_x+c_0H_{0}\right)$ where $\lambda_0,b_0,c_0$ 
are real parameters. When an incident electromagnetic radiation with frequency $\omega$ in the vicinity of one of these two resonances enters the bus 
waveguide, in either direction, it will primarily excite the closer mode in frequency without being (to a good approximation) affected by the presence 
of the other resonance. Below we consider the case $\omega\approx \omega_s$ where maximum non-reciprocity is observed. Therefore 
we will assume that the incident wave is coupled directly only with the symmetric mode. 

Under these assumptions, the temporal evolution of the symmetric ($a_s$) and antisymmetric ($a_a$) modal amplitudes is 
described by the following equations
\begin{align}
\frac{\mathrm{d}a_{s}}{dt}= & \imath\omega_{s}^{(0)}a_{s}-\frac{1}{\tau}a_{s}-\lambda^{*}a_{a}+\kappa_{1}
S_{1}^{in}+\kappa_{2}S_{2}^{in}\nonumber \\
\frac{\mathrm{d}a_{a}}{dt}= & \imath\omega_{a}^{(0)}a_{a}+\lambda a_{s}\label{eq:key}\\
S_{-}^{out}= &S_{2}^{in}-\kappa_2^* a_{s};\quad S_{+}^{out}= S_{1}^{in}-\kappa_1^* a_{s}\nonumber 
\end{align}
where $\frac{1}{\tau}=\frac{1}{\tau_{-}}+\frac{1}{\tau_{+}}$ is the radiative coupling of the symmetric mode to a left-going 
(${1\over \tau_{-}}$) or a right-going (${1\over \tau_{+}}$) output wave, and $\left\{ \kappa_{1},\kappa_{2}\right\}$ indicate 
the coupling constants between the symmetric mode and the incoming or outgoing waves. We have that $|\kappa_1|^2={2
\over \tau_{+}}$ and $|\kappa_2|^2={2\over \tau_{-}}$. The modal amplitudes are normalized in such a way that $\left|a_{s}
\right|^{2}$ ($\left|a_{a}\right|^{2}$) correspond to the energy stored at the specific mode, while $\left|S_{1}^{in} \right|^{2}$ 
and $\left|S_{2}^{in}\right|^{2}$ ($\left|S_{-}^{out}\right|^{2}$ and $\left|S_{+}^{out} \right|^{2}$) are the powers carried by 
incoming (outgoing) waves from (to) two different directions of the bus waveguide.

The forward $T_F\equiv \frac{\left|S_{+}^{out}\right|^{2}}{\left|S_{1}^{in}\right|^{2}}$ and backward 
$T_B\equiv \frac{\left|S_{-}^{out}\right|^{2}}{\left|S_{2}^{in}\right|^{2}}$ transmittance for a left $S_{1}^{in}\propto 
e^{\imath\omega t}$ and right $S_{2}^{in}\propto e^{\imath\omega t}$ incident monochromatic field can be 
calculated from Eq. (\ref{eq:key}) by imposing the appropriate boundary conditions $S_{2}^{in}=0$ and $S_{1}^{in}=0$ 
respectively. We obtain that
\begin{equation}
T_{\rm{F/B}}\left(\omega\right)=
\left|
\frac{i\left(\omega-\omega_s^{(0)}-\frac{\left|\lambda_{\rm F/B}\right|^{2}}{(\omega-\omega_a^{(0)})}\right) \mp \Delta\epsilon}
{i\left(\omega-\omega_s^{(0)}-\frac{\left|\lambda_{\rm F/B}\right|^{2}}{\left(\omega-\omega_{a}^{(0)}\right)}\right)+\left(\frac{1}{\tau}\right)}
\right|^2
\label{TF}
\end{equation}
where $\Delta\varepsilon=\frac{1}{\tau_{+}}-\frac{1}{\tau_{-}}\neq 0$ due to gyrotropy and $\lambda_{F/B}$ is the coupling between
$\omega_{s}^{(0)}$ and $\omega_a^{(0)}$ for forward and backward propagation. 

From Fig. \ref{fig3} we observe that the maximum $NR$ occurs at the resonance frequency $\omega_s^B$ of the BWD propagation which 
can be estimated from Eq. (\ref{TF}) to be $\omega_{s}^{\rm B}= \omega_{0}-\sqrt{\Omega^2-(\rho\gamma)^2}$. The dependence 
of $\omega_s^B$ on $H_0$ allows us to reconfigure the position of maximum non-reciprocity. The modified coupling $\Omega\equiv
\sqrt{\Omega_0^{2}+\left|\lambda_{\rm B}\right|^{2}}$ is a result of the external magnetic field which now also acts at the substrate between 
the two cavities and the presence of the incident wave in the bus waveguide. It allows us to estimate the gain and loss parameter $\tp{\gamma_{\cal 
{\tilde P}T}}=\Omega/\rho$ for which we have an EP singularity for the isolated system with the uniform magnetic field (see Fig. \ref{fig2}).

These theoretical results compare nicely with the COMSOL simulations in Fig. \ref{fig3} in the domain of $\omega\approx\omega_s^B$. 
A non-linear least square fit has been used in order to fit Eq. (\ref{TF}) to the data for $T_{\rm B}$. The parameters that we have obtained are 
$\Delta \epsilon\approx -0.0075$, ${1\over \tau}\approx 0.05215$, $\eta\approx 4.9\times 10^{-3}$ (all measured in nsec$^{-1}$) and 
$|\lambda_B|^2\approx 0.111$ nsec$^{-2}$. All these parameters, apart from $|\lambda_F|^2$, have been kept fixed for the forward 
transmission $T_F$, see Eq. (\ref{TF}). The fitting value of $T_F$ indicated that $|\lambda_F|^2\approx |\lambda_B|^2 $ nsec$^{-2}$. Finally, using 
Eqs. (\ref{TF}) together with  Eq. (\ref{NR}) we have calculated $NR$ versus $\gamma$. These theoretical results are shown in Fig. 
{\ref{fig4}b together with the simulations of COMSOL.   

In order to enhance our understanding of the origin of the giant nonreciprocal effect we have further approximated $NR$ at $\omega=
\omega_s^B$. Guided by the numerics, which indicates that $T_F(\omega_s^B)\sim {\cal O}(1)$ in this frequency domain, 
we have assumed that  $\log_{10}T_{\mathrm{F}}\left(\omega_{s}^{\mathrm{B}}\right)$ is negligible when compared to 
$\log_{10}T_{\rm B}\left(\omega_{s}^{\rm B}\right)$. Therefore $NR(\gamma)\approx 10|\log_{10} T_B(\omega_s^{\rm B})|$. 
This approximation leads us to the following expression up to leading order in $\eta,\Delta \epsilon$ and $\epsilon\equiv1/(2\tau)$
\cite{note2}:
\begin{equation}
\label{NRapprox}
\mathrm{NR}=\left\{
\begin{array}{cc}
20\log_{10}
\frac{1+\frac{\varepsilon}{\eta}\left(1+\frac{\sqrt{\beta}}{\sqrt{1+\beta}}\right)}
{1+\frac{\Delta\varepsilon}{2\eta}\left(1+\frac{\sqrt{\beta}}{\sqrt{1+\beta}}\right)}
; & 0<\gamma<\tp{{\gamma_{\cal {\tilde P}T}^{0}}} \\
10\log_{10}
\frac{\left(\eta+\varepsilon\right)^2+\beta\eta(\eta+2\varepsilon)}
{\left(\eta+\Delta\varepsilon/2\right)^{2}+\beta\eta \left(\eta+\Delta\varepsilon\right)}
; & \tp{\gamma_{\cal {\tilde P}T}^{0}}<\gamma<\tp{\gamma_{\cal {\tilde P}T}}
\end{array}\right.
\end{equation}
 where $\beta\equiv\frac{\Omega_0^{2}-(\rho\gamma)^{2}}{\left|\lambda_{\rm B}\right|^{2}}$.

A further analysis of Eq. (\ref{NRapprox}), indicates that when $\frac{\Delta\varepsilon}{2\eta}<\min\left\{ -\frac{\Omega}{\Omega_0
+\Omega},\,-\frac{\varepsilon}{2\varepsilon+\eta}\right\}$, then $NR(\gamma)$ has a single maximum in the exact phase i.e. $0\leq 
\gamma\leq \tp{\gamma_{\cal \tilde{P}T}}(H_{0})$ which occur at some critical value $\gamma=\gamma_{\rm NR}$. In case $\frac{\Delta
\varepsilon}{2\eta}<-1$, we have $\gamma_{\rm NR}=\tp{\gamma_{\cal {\tilde P}T}^{0}}$ while for $-1<\frac{\Delta\varepsilon}{2\eta}<\min
\left\{ -\frac{\Omega}{\Omega_0+\Omega},\,-\frac{\varepsilon}{2\varepsilon+\eta}\right\}$ we have $\gamma_{\rm NR}=\sqrt{
(\tp{\gamma_{\cal {\tilde P}T}^{0}})^{2}-\frac{\left|\lambda_{\rm B}/\rho\right|^{2}}{\left(\frac{\Delta\varepsilon}{2\eta}/\left(1+\frac{\Delta\varepsilon}
{2\eta}\right)\right)^{2}-1}}$. Thus we conclude that the existence and position of $\gamma_{\rm NR}$ is strongly dictated by 
$ \tp{\gamma_{\cal {\tilde P}T}^{0}}$ and $|\lambda_B|^2$, i.e., this giant non-reciprocal behavior is a consequence of an interplay between 
the EP degeneracy and the interaction of fields within the gyrotropic substrate.


\section{Lumped Circuit Analysis}

The EP-induced giant non-reciprocity can be further analyzed utilizing an electronic circuit 
analog that maintains the essence of the original physics while also allowing a significantly simplified path toward both analytic and 
numeric analysis. The circuit, shown in \tp{Fig.~\ref{fig5}(a)}, reduces the parallel microstrip resonators to a pair of $RLC$ resonators 
capacitively coupled to points separated by a distance $d$ along an ideal TEM transmission line. The inter-resonator coupling through 
the gyrotropically active substrate is incorporated as a mutual inductance $M$ in parallel with an ideal gyration $G$ such that the 
inductor currents are related to the voltages by
\begin{equation}
\left(\begin{array}{c}
I_{1}\\
I_{2}
\end{array}\right)=\frac{1}{i\omega}\begin{bmatrix}
  L & M\\
  M & L\\
\end{bmatrix}^{-1}\left(\begin{array}{c}
V_{1}\\
V_{2}
\end{array}\right)+\begin{bmatrix}
  0 & G\\
  -G & 0\\
\end{bmatrix}\left(\begin{array}{c}
V_{1}\\
V_{2}
\end{array}\right)
\label{gyro}
\end{equation}
The gain and loss, along with the small inherent loss $\eta$ defined earlier, are implemented by negative and positive parallel resistances of slightly different magnitude.

In the frequency domain, Kirchoff's Laws for this circuit are easily expressed, though transcendental due to the trigonometric wave components in the center transmission 
line section. All seven element of the circuit ($L$, $C$, $R_1$, $R_2$, $M$, $G$, $C_c$, and $d$)  represent essential features of the original structure that can contribute 
to the enhancement of the transmission nonreciprocity. Note that $G$ plays a similar role as the static magnetic field $H_0$ in the gyrotropic substrate of the microstrip device and is the key circuit element responsible for nonreciprocity.

\begin{figure*}
\hspace*{-0cm}\includegraphics*[width=16cm]{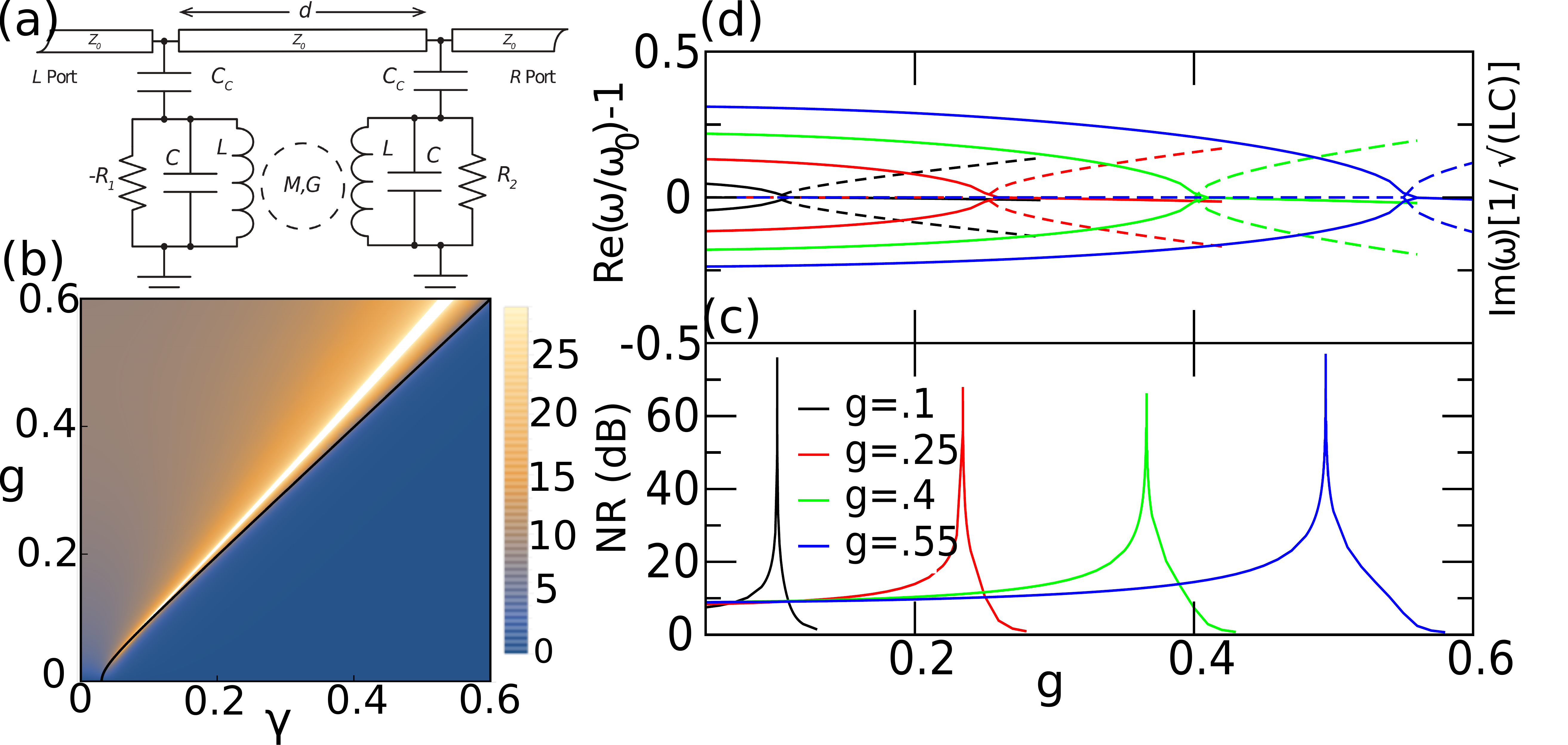}
\caption{Exploration of the gain/loss $\gamma=\frac{1}{2}(R_1^{-1}+R_2^{-1})\sqrt{L/C}$ and gyration strength $g=G\sqrt{L/C}$ parameter space of the lumped circuit 
model shown in (a). In (b) we plot the nonreciprocity NR as intensity (high values of NR correspond to bright areas while low values of NR to 
dark areas) in the map. Due to the limitation of the resolution, the narrow peaks representing high NR ($>$ 30 dB) in Fig. 5 (c) are not resolved by the color bar. (c) We show some indicative "cuts" from the density map at several gyration strengths \tp{(shown in (b))}
for $Z_0\sqrt{C/L}=0.82$, $kd\approx\pi$ at the $LC$ resonant frequency, $C_c/C=0.3$, $M/L=0.03$, and $\eta=\frac{1}{2}(R_2^{-1}-R_1^{-1})\sqrt{L/C}=0.03$ for the 
intrinsic loss. (d) Shows the corresponding real and imaginary parts of the balanced, isolated ($\eta=C_c=0$) dimer mode frequencies illustrating 
relation of the exceptional points to the singularities of the giant non-reciprocity. The solid line through the NR density plot shows the position of the isolated system exceptional 
point, slightly beyond the singularity.}
\label{fig5}
\end{figure*}

The main graphs of Fig.~\ref{fig5}\tp{(b)-(d)} illustrate numerical results exploring the NR with gain/loss and gyration strengths, $\gamma=\frac{1}{2}(R_1^{-1}+R_2^{-1})\sqrt{L/C}$ 
respectively, to a detail that is computationally expensive in the COMSOL simulation, and somewhat abstract in the theoretical analysis. The NR density plot shown in Fig.~\ref{fig5}(b) is 
separated into two regions by the black solid line representing the position of the isolated exceptional point, with the exact ${\cal {\tilde P}T}$ phase above and the broken phase 
below. The singular NR is seen as the bright swath {\it within the unbroken region} just above \cite{note0}. Figure~\ref{fig5}(c) show cuts of the NR at 
several fixed values of the gyration strength $g$ (below) along with the corresponding isolated dimer eigenfrequencies (above). Note again that the maximum NR occurs 
below the isolated exceptional points. The similarities with Fig. \ref{fig4} associated with the photonic structure is striking, thus indicating the shared NR mechanism.
Specifically for $\gamma=0$ we again observe a moderate non-reciprocal behavior which is dramatically enhanced at $\gamma$-values close to $\gamma_{\cal {\tilde P}T}$.
This can be better appreciated by analyzing the parametric evolution of the eigenfrequencies of the isolated circuit. The isolated system in this electronic analog includes all 
of the effects of the resonator coupling, such as the gyration, fulfilling the inequality expressed in Eq. \ref{NRapprox}.

\section{Discussion} 

We have shown that the flexibility introduced by the ${\cal {\tilde P}T}$ properties of the photonic resonator dimer dramatically enhance the strength -- and hence the bandwidth -- 
of the singular nonreciprocity. We have observed this over certain ranges of the system parameters. At the same time we have demonstrated that these results apply equally 
well in the case of lumped circuitry.

This universal nature of the giant non-reciprocal response near the EP calls for an intuitive explanation. First we have to realize that the structure constitutes an effective ring 
since the two cavities are directly coupled to one-another while at the same time they are coupled indirectly via the bus waveguide. At the EP the two supermodes of the effective 
ring structure are degenerate having a definite chirality \cite{lan1}. The presence of the magnetic field breaks the spectral degeneracy, while weakly preserving the (common) chiral nature 
of the modes. As a result the two modes are coupled differently with a left and a right incident wave. Assuming, for example in the electronic set up of Fig.~\ref{fig5}(a), that the chirality of the modes is clockwise (CW) 
we conclude that due to phase matching such a mode will be coupled only to a left incident wave but not to a right incident one. Accordingly, the left incident wave will excite 
the CW supermode while at the same time can exploit a direct optical path associated with a transmission via a direct process between the incident and transmitted channels. 
These optical paths can interfere destructively at the output channel (depending on the propagation phase associated with the length of the bus waveguide and the gyrotropy) 
leading to a Fano effect and consequently to a (near) zero transmittance. An important condition here is that the internal losses of the cavities are small so that the two interfering 
waves have the same amplitudes. On the other hand, a right incident wave, because of phase mismatch, does not couple to the CW chiral supermode of the effective ring. As a 
result it does not experience the internal losses inside the cavity and consequently the (direct) transmission is high.

The electronic circuit that we proposed can be realized experimentally using existing MOSFET technologies. Such reconfigurable circuitry (due to on-the-fly manipulation 
of gain and loss) would be useful in the realization of RF circulators and isolators. Moreover in the optical domain where the magneto-optical effects are very weak, the wave propagation can be masked by unwanted losses associated with the materials used as a means to realize non-reciprocal propagation. Our scheme -- with the manipulated gain 
and loss -- would resolve some of the above mentioned issues and help restore a strong non-reciprocal signal.


\section{Conclusions } 

We have theoretically defined the conditions for which a non-Hermitian structure with antilinear symmetry
can lead to giant non-reciprocal transport: the system has to operate in the stable domain and in the vicinity of an EP singularity
which amplify the effects of gyrotropy. The non-reciprocal frequency domain is reconfigurable, albeit is narrow-band. We have 
demonstrated the validity of the theoretical predictions in the microwave domain where we have found non-reciprocal transmission 
which is higher than 90dB. We have further confirm the generality of our results utilizing a user-friendly framework of lump circuits.
It will be interesting to extend this study and investigate giant non-reciprocal transport in acoustic or matter-wave systems where 
amplification and attenuation mechanisms can be easily controlled and used to realize EP degeneracies \cite{acousticPT,BECPT} while an effective magnetic field 
can be introduced via time-varying potentials \cite{alu,bec}.

{\it Acknowledgements -}This work was partly sponsored by AFOSR MURI Grant No. FA9550-14-1-0037, and by NSF Grant No. 
DMR-1306984.



\begin{thebibliography}{99}

\bibitem{K80} T. Kato, {\it Perturbation Theory of Linear Operators}, Springer Classics in Mathematics (Springer, NY), (1980).

\bibitem{B04} M. V. Berry, Cechoslov. J. Phys. {\bf 54}, 1-39 (2004)

\bibitem{BB98} C. M. Bender, S. Boettcher, Phys. Rev. Lett. {\bf 80}, 5243 (1998); C. M. Bender, Rep. Prog. Phys. {\bf 70}, 947 (2007).

\bibitem{M11} N. Moiseyev, {\it Non-Hermitian Quantum Mechanics}, Cambridge Univ. Press (2011).

\bibitem{H12} W. D. Heiss, J. Phys. Math. Theor. {\bf 45}, 444016 (2012).

\bibitem{Guo}A. Guo, {\it et al.}, Phys. Rev. Lett. {\bf 103}, 093902 (2009).

\bibitem{LREKCC11} Z. Lin, H. Ramezani, T. Eichelkraut, T. Kottos, H. Cao, D. N. Christodoulides,
Phys. Rev. Lett. {\bf 106}, 213901 (2011).

\bibitem{FXFLOACS13} L. Feng {\it et al.}, Science {\bf 333}, 729 (2011).

\bibitem{GSDMVASC09}  A. Regensburger, C. Bersch, M.A. Miri, G. Onishchukov, D. N. Christodoulides, U. Peschel,
Nature {\bf 488}, 167 (2012).

\bibitem{HMHCK14} H. Hodaei {\it et al.}, Science {\bf 346}, 975 (2014); L. Feng {\it et al.}, Science {\bf 346}, 972 (2014).

\bibitem{PORYLMBNY14} B. Peng, S. K. Ozdemir, S. Rotter, H. Yilmaz, M. Liertzer, F. Monifi, C.M. Bender, F. Nori, L. Yang, 
Science {\bf 346}, 328 (2014); M. Chitsazi, S. Factor, J. Schindler, H. Ramezani, F. M. Ellis, and T. Kottos, Phys. Rev. A {\bf 89}, 043842 (2014);
M. Brandstetter, M. Liertzer, C. Deutsch, P. Klang, J. Sch\"oberl, H. E. T\"ureci, G. Strasser, K. Unterrainer, and S. Rotter,
Nat. Comms. {\bf 5} (2014).

\bibitem{lan1} B. Peng, S. K. \"Ozdemir, M. Liertzer, W. Chen, J. Kramer, H. Yilmaz, J. Wiersig, S. Rotter, L. Yang, PNAS {\bf 113}, 6845 (2016).

\bibitem{lan2} J. Wiersig, Phys. Rev. A {\bf 93}, 033809 (2016); Z-P Liu, J. Zhang, S. K. \"Ozdemir, B. Peng, H. Jing, X-Y L\"u, C-W Li, L Yang, 
F. Nori, Y-x Liu, arXiv:1510.05249.

\bibitem{note0} In Ref. \cite{RLKKKV12} we have studied non-reciprocal transport from non-Hermitian layered structures with magnetic
layers when the system is in the broken phase (see for example Fig. 2(a) of Ref. \cite{RLKKKV12}). An important requirement for non-reciprocity 
in this case is a broken space inversion symmetry. This was achieved with the use of misaligned birefringent layers. In contrast, the current study 
singles out the importance of EPs and demonstrates that  {\it giant} non-reciprocity occurs in the exact (stable) phase 
provided that the system is in the proximity of an EP.

\bibitem{RLKKKV12} H. Ramezani {\it et al.}, Optics Express {\bf 20},  26200 (2012).

\bibitem{MGCM08} K. G. Makris {\it et al.}, Phys. Rev. Lett. {\bf 100}, 103904 (2008).

\bibitem{RMGCSK10} C. E. R\"uter {\it et al.}, Nat. Phys. {\bf 6}, 192 (2010).

\bibitem{Konotop} F.Kh. Abdullaev, V.V. Konotop, M.Ogren, and M.P. Soerensen, Opt.Lett. {\bf 36}, 4566 (2011).

\bibitem{CJHYWJLWX14} L. Chang {\it et al.}, Nat. Phot. {\bf 8}, 524 (2014).

\bibitem{POLMGLFNBY14} B. Peng {\it et al.}, Nat. Phys. {\bf 10}, 394 (2014).


\bibitem{L09a}S. Longhi, Phys. Rev. Lett. {\bf 103}, 123601 (2009).

\bibitem{ZCFK10} M. C. Zheng {\it et. al},  Phys. Rev. A {\bf 82}, 010103 (2010).

\bibitem{RKGC10}H. Ramezani, T. Kottos, R. El-Ganainy, and D. N. Christodoulides, Phys. Rev. A {\bf 82}, 043803 (2010).

\bibitem{BFBRCEK13}N. Bender, S. Factor, J. D. Bodyfelt, H. Ramezani, D. N. Christodoulides, F. M. Ellis, 
and T. Kottos, Phys. Rev. Lett, {\bf 110}, 234101 (2013).

\bibitem{NBRMCKF14}F. Nazari, N Bender, H Ramezani, MK Moravvej-Farshi, DN Christodoulides, T Kottos, Optics express {\bf 22}, 9574 (2014).

\bibitem{LTEK16}\tp{H. Li, R Thomas, F. M. Ellis, T Kottos, New J. Phys. {\bf 18}, 075010 (2016)}

\bibitem{FHWWCYL11} \tp{Y. Fan, J. Han, Z. Wei, C. Wu, Y. Cao, X. Yu, H. Li, Appl. Phys. Lett. {\bf 98}, 151903 (2011).}

\bibitem{SDBB94}\tp{M. Scalora, J. P. Dowling, C. M. Bowden, M. J. Bloemer, J. Appl. Phys. {\bf 76}, 2023 (1994).}

\bibitem{FSK05}\tp{M. W. Feise, I. V. Shadrivov, Y. S. Kivshar, Phys. Rev. E {\bf 71}, 037602 (2005).}

\bibitem{YF09}\tp{Z. Yu, S. Fan, Nat. Photonics {\bf 3}, 91 (2009)}

\bibitem{MDE14}\tp{A. M. Mahmoud, A. R. Davoyan and N. Engheta, Nat. Communications {\bf 6}, 8359 (2015).}

\bibitem{DBB12}\tp{D. Dai, J. Bauters and J. E. Bowers, Light: Science and Applications {\bf 1}, 1 (2012).}

\bibitem{Shvets_14}S. Mousavi, S. Hossein, A. B. Khanikaev, J. Allen, M. Allen, G. Shvets, Phys. Rev. Lett. {\bf 112}, 117402 (2014).

\bibitem{Song_Wu10}J. H. Leach, H. Liu, V. Avrutin, E. Rowe, U. Ozgur, H. Morkoc, Y-Y Song, M. Wu, Journal of Applied Physics {\bf 108}, (2010).

\bibitem{Fred_12}J. Schindler, Z. Lin, J. M. Lee, H. Ramezani, F. M. Ellis, T. Kottos, J. Phys. A: Math. Theor. {\bf 45},
444029 (2012).

\bibitem{Matolak_12}S. Laha, S. Kaya, A. Kodi, D. Matolak, ``Double gate MOSFET based efficient wide band tunable power amplifiers",
{\it Wireless and Microwave Technology Conference (WAMICON), 2012 IEEE 13th Annual}, pages 1-4 (2012) 
[doi=10.1109/WAMICON.2012.6208460].

\bibitem{comsol} Comsol Multiphysics, Comsol {\bf 5} (2015). Further details can be found in the book "Multiphysics Modeling With Finite Element Methods" by W. Zimmerman

\bibitem{note1} There are, in practice, intrinsic losses which result to a trivial identical imaginary part for the eigen-modes. These 
losses can be scaled out from the equations that describe the system and considered at the final stage by addition to the bare 
eigen-frequencies $\omega_0$ an imaginary part i.e. $\omega_0\rightarrow \omega_0+i\eta$.

\bibitem{H84} H. Haus, Waves and Fields in Optoelectronics; Prentice-Hall: Englewood Cliffs, NJ, (1984)

\bibitem{note2} We have considered that the intrinsic loss is the same for both symmetric and antisymmetric modes i.e 
$\eta_{s}=\eta_{a}=\eta$. 

\bibitem{acousticPT} R. Fleury, D. L. Sounas, and A. Al\'u, Nat. Comms. {\bf 6},  5905 (2015); 
C. Shi, M. Dubois, Y. Chen, L. Cheng, H. Ramezani, Y. Wang,	X. Zhang, Nat. Comms. {\bf 7}, 11110 (2016). 

\bibitem{BECPT}F. Single, H. Cartarius, G. Wunner, J. Main, Phys. Rev. A {\bf 90}, 042123 (2014); K. Ding, G. Ma, M. Xiao, Z. Q. Zhang, 
C. T. Chan, Phys. Rev. X {\bf 6}, 021007 (2016)

\bibitem{alu}R. Fleury, D. L. Sounas, C. F. Sieck, M. R. Haberman, A. Al\'u, Science  {\bf 343}, 516 (2014).

\bibitem{bec}J. Struck, C. \"Olschl\"ager, M. Weinberg, P. Hauke, J. Simonet, A. Eckardt, M. Lewenstein, K. Sengstock, and P. Windpassinger, 
Phys. Rev. Lett. {\bf 108}, 225304 (2012); K. Jim\'enez-Garc\'ia, L. J. LeBlanc, R. A. Williams, M. C. Beeler, A. R. Perry, and I. B. Spielman, Phys. Rev. 
Lett. 108, 225303 (2012).


 
\end{thebibliography}
\end{document}